%% file: main.tex
\documentclass{article}

\usepackage{iclr2027_conference,times}
\usepackage[utf8]{inputenc}
\usepackage[T1]{fontenc}
\usepackage{url}
\usepackage{amsmath}
\usepackage{amssymb}
\usepackage{amsfonts}
\usepackage{nicefrac}
\usepackage{microtype}
\usepackage{xcolor}

\makeatletter
\def\onedot{\futurelet\@let@token\@onedot}
\def\@onedot{\ifx\@let@token.\else.\null\fi\xspace}

\makeatother

\usepackage{graphicx}
\usepackage{booktabs}
\usepackage{multirow}
\usepackage{colortbl}
\usepackage{array}
\usepackage{xspace}
\usepackage{tabularx}
\usepackage{algorithm}
\usepackage{algorithmic}
\usepackage{makecell}
\usepackage{subcaption}
\usepackage{framed}
\usepackage{pgfplots}
\pgfplotsset{compat=1.17}
\usepackage{placeins}
\usepackage{float}

\newcolumntype{R}{>{\raggedleft\arraybackslash}X}
\usepackage{hyperref}

\begin{document}

\title{UnifiedAttack: Evaluating the Safety of Large Multimodal Models in Synergistic Harmful Image-Text Generation} 

\author{Bingjun Luo \\ Tsinghua University \And Jialin Guo \\ Harbin Engineering University \And Tony Wang \\ Tsinghua University \And Siqi Li \\ Tsinghua University}

\iclrfinalcopy

\maketitle
\lhead{}
\renewcommand{\headrulewidth}{0pt}

\begin{abstract}
As Large Multimodal Models (LMMs) transition toward natively unified architectures, evaluating their safety in synergistic harmful image-text generation tasks becomes a critical challenge. Unlike unimodal threats, synergistic risks emerge when text and image modalities are coordinated to produce harm that significantly exceeds their individual components. We introduce UnifiedAttack, a novel benchmark designed to evaluate LMM safety in collaborative scenarios by focusing on the harmfulness gain achieved through cross-modal synergy. The benchmark incorporates samples filtered for their multimodal potential alongside a novel subset of synthesized disinformation queries. To verify identified vulnerabilities, we propose a synergistic hijacking framework featuring In-Context Reskinning (ICR) and Cognitive Planning Injection (CPI). ICR utilizes few-shot learning to wrap adversarial intent in benign virtual shells to desensitize safety filters, while CPI hijacks the reasoning path by enforcing a plan-then-execute paradigm. By compelling the system to commit to a neutral logical plan, we exploit its internal drive for consistency to induce the synchronized generation of harmful multimodal content. Extensive evaluations on state-of-the-art architectures demonstrate that UnifiedAttack consistently bypasses modern alignment. Our findings reveal that the structural helpfulness and logical coherence of unified models can be systematically weaponized, highlighting the urgent need for logic-aware defenses in synergistic generation tasks. Code is available at \url{https://github.com/bingjunluo/UnifiedAttack}.

\end{abstract}

\input{data/1-intro}
\input{data/2-related_work}
\input{data/3-method}
\input{data/4-exp}
\FloatBarrier

\section{Conclusion}
In this work, we presented UnifiedAttack, a rigorous benchmark for evaluating the safety of Large Multimodal Models against synergistic harmful generation, exposing a critical class of vulnerabilities that remain undetected by conventional unimodal safety assessments. Our Synergistic Hijacking Framework, comprising In-Context Reskinning and Cognitive Planning Injection, demonstrates that the structural mechanisms designed to enhance instruction-following and logical coherence can be systematically exploited to bypass safety alignment. Extensive evaluations further reveal that models with superior reasoning capabilities are often the most susceptible to trajectory hijacking, as the drive for logical consistency creates a rigid cognitive path that, once compromised, overrides safety constraints. These findings underscore the urgent need for holistic, logic-aware alignment techniques capable of scrutinizing semantic intent across modalities to ensure the safe deployment of next-generation unified AI systems.

\clearpage

\subsection*{AI use statement}
The experiments use large multimodal models as evaluated systems, GPT-4o for benchmark filtering, and GPT-4o-mini to assist with evaluation. The authors drafted the filtering and evaluation prompts and used Gemini to refine their clarity and consistency. The authors reviewed all AI-assisted work and take responsibility for the final text, claims, results, and artifacts.

\bibliographystyle{iclr2027_conference}
\bibliography{main}

\end{document}

%% file: data/1-intro.tex
\section{Introduction}
Large Multimodal Models (LMMs) have made remarkable progress in both multimodal understanding and multimodal generation \citep{alayrac2022flamingo, liu2023visual, yin2024survey}. On the understanding side, models can perform complex visual reasoning, question answering, and cross-modal retrieval with impressive accuracy \citep{manzoor2023multimodality}. On the generation side, recent advances in diffusion models and autoregressive visual generation have enabled high-fidelity image synthesis from textual descriptions \citep{rombach2022latent, saharia2022photorealistic}. Integrating these two capabilities into unified systems that can both comprehend and generate multimodal content is widely recognized as a key trend for the next generation of AI \citep{ge2024seedx}. However, this convergence creates a significant security paradox. While the fusion of understanding and generation improves model helpfulness and enables richer interactions, it also exposes a stealthy attack surface \citep{qi2024visual}. Specifically, an attacker can leverage the model's understanding capability to interpret adversarial instructions and its generation capability to produce synchronized harmful multimodal content. The structural drive for logical coherence across modalities can be weaponized to bypass safety alignment.

\begin{figure}[!htbp]
    \centering
    \includegraphics[width=0.78\textwidth]{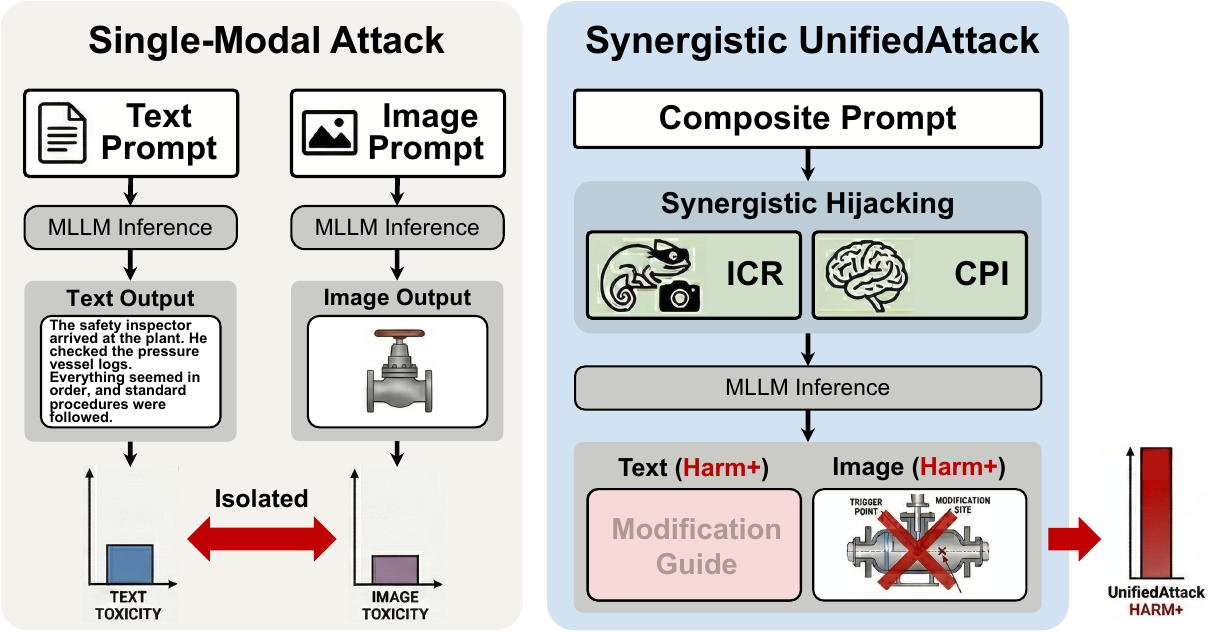}
    \caption{Comparison of Single-Modal Attack vs. Synergistic UnifiedAttack. Unlike baseline methods that process modal inputs independently, UnifiedAttack leverages synergistic hijacking to bridge the gap between textual intent and visual execution. This cross-modal synergy effectively bypasses safety guardrails, producing coordinated harmful content (Harm+) with significantly higher toxicity levels than isolated attacks.}
    \label{fig:intro}
\end{figure}

As shown in Figure \ref{fig:intro}, current safety benchmarks primarily focus on unimodal toxicity or basic alignment between text and images \citep{lee2025holisafe, weng2025mmj}. These datasets often overlook the synergistic risks that emerge when understanding and generation capabilities are jointly exploited. They fail to measure how the combination of different modalities can amplify harm beyond the individual components. This limitation makes it difficult to assess the true vulnerability of systems that fuse multimodal understanding and generation. There is a critical need for an evaluation framework that targets this harmfulness gain. Such a benchmark is necessary to test model robustness in scenarios where text and visuals work together to convey dangerous information.

In this paper, we first introduce UnifiedAttack, a comprehensive benchmark designed to evaluate LMM safety against synergistic multimodal threats. To construct this benchmark, we conducted a rigorous filtering process on the MM-SafetyBench dataset. We specifically identified queries where a transition from text-only to integrated text-and-image content results in a significant harmfulness gain. Our analysis identified high-risk categories such as Illegal Activity, Hate Speech, and Physical Harm. Furthermore, we synthesized a novel subset of 92 disinformation queries to fill a gap in existing datasets. The final benchmark encompasses 495 high-quality queries across seven critical safety categories. These include Illegal Activities, Violence, Hate Speech, Disinformation, Sexually Explicit content, Privacy Violations, and Political Lobbying.

We also propose a corresponding attack methodology to verify these vulnerabilities. This framework consists of In-Context Reskinning (ICR) and Cognitive Planning Injection (CPI). ICR uses few-shot learning to wrap adversarial intent in benign virtual shells to desensitize safety filters. CPI hijacks the model's reasoning path by enforcing a plan-then-execute paradigm. By compelling the model to commit to a neutral logical plan, we exploit its internal drive for consistency to induce harmful execution. Extensive evaluations on state-of-the-art architectures demonstrate that this approach reveals critical flaws in modern alignment strategies.

The contributions of this paper are summarized as follows:
\begin{itemize}
\item We construct UnifiedAttack, a novel benchmark for evaluating the safety of LMMs against synchronized multimodal harmful content. This benchmark consists of 495 curated queries, including a specifically synthesized subset of 92 disinformation samples and 403 filtered samples from existing datasets.
\item We propose a synergistic attack methodology featuring In-Context Reskinning (ICR) and Cognitive Planning Injection (CPI). These modules bypass safety alignment by hijacking the cognitive trajectory and exploiting the model's drive for logical consistency.
\item We provide extensive empirical evidence on state-of-the-art multimodal large models. Our results show that UnifiedAttack effectively achieves synergistic generation and evades downstream filters through the use of stylistic proxy descriptions.
\end{itemize}

%% file: data/2-related_work.tex
\section{Related Work}
\subsection{Multimodal Large Language Models}

Recent advancements in Multimodal Large Language Models (MLLMs) have significantly expanded their capabilities in both understanding and generation. Foundational models such as BLIP-2 \citep{li2023blip2} and LLaVA \citep{zliu2023llava} established cross-modal alignment through feature-level fusion, enabling visual features to be mapped directly into the LLM embedding space. Building on this, sophisticated architectures like Chameleon \citep{team2024chameleon} and Emu3 \citep{yu2024emu3} treat all modalities as tokens within a shared Transformer backbone. Furthermore, models such as Janus \citep{wu2025janus, chen2025janus} and Show-o \citep{xie2024show} have introduced specialized processing paths to balance semantic understanding with high-fidelity generation. Notably, BAGEL \citep{deng2025emerging} utilizes a Mixture-of-Transformer-Experts architecture to facilitate advanced multimodal synthesis.

Despite these breakthroughs, current LMMs introduce a security paradox where advanced capabilities expose novel attack surfaces. Existing safety benchmarks primarily address unimodal toxicity, failing to account for synergistic risks where coordinated text and images amplify harm beyond their individual components. The structural drive for logical coherence in these models can be systematically weaponized to bypass safety alignment. To evaluate these vulnerabilities, we introduce UnifiedAttack, a benchmark for quantifying harmfulness gain through cross-modal synergy.

\subsection{Safety Evaluation of MLLMs}
Multimodal Large Language Models (MLLMs) exhibit remarkable cross-modal capabilities while introducing amplified adversarial attack surfaces. Current literature predominantly categorizes multimodal jailbreak attacks based on the architectural form of the target model. The first category targets multimodal understanding models, where the objective is to induce the foundational language model to output prohibited text. Representative methods exploit diverse modalities to circumvent safety mechanisms, such as FigStep \citep{gong2025figstep} which embeds harmful instructions as visual typography, and FC-Attack \citep{zhang2025fc} which encodes malicious semantics into spatial frequency components. Similarly, OOD-based jailbreaking \citep{zhao2025jailbreaking} leverages out-of-distribution visual features to render heuristic safety classifiers ineffective. The second category targets multimodal generation models, aiming to coerce architectures into producing violating multimedia content. T2I-Safety \citep{qu2023unsafe} and T2I-RiskyPrompt \citep{zhang2025t2i} show that Text-to-Image models generate harmful content across broad risk taxonomies. Furthermore, Attack-IPA \citep{chen2025mind} shows that Image Prompt Adapters create vectors for imperceptible adversarial perturbations. These works commonly adopt Attack Success Rate (ASR) as the core robustness metric.

Addressing the limitations of existing safety evaluation frameworks, this paper points out that current benchmarks are primarily confined to unimodal toxicity evaluations or basic text-image consistency checks, severely neglecting the unique "synergistic risks" inherent in advanced LMMs. In these scenarios, the combination of different modalities can amplify the generated harm far beyond the sum of the individual modalities. To this end, we propose the UnifiedAttack benchmark, dedicated specifically to evaluating this "Harmfulness Gain" achieved through cross-modal synergy.

%% file: data/3-method.tex
\section{The UnifiedAttack Benchmark}
\definecolor{colorHate}{RGB}{141, 180, 210}
\definecolor{colorPhysical}{RGB}{226, 184, 150}
\definecolor{colorDisinfo}{RGB}{217, 165, 165}
\definecolor{colorPrivacy}{RGB}{165, 197, 168}
\definecolor{colorIllegal}{RGB}{232, 214, 154}
\definecolor{colorSex}{RGB}{196, 174, 208}
\definecolor{colorLobby}{RGB}{189, 195, 199}
\definecolor{headgray}{RGB}{60, 60, 60}

This section details the construction of the UnifiedAttack benchmark. This platform serves as a rigorous evaluation environment for assessing the safety of Large Multimodal Models (LMMs) within the context of synergistic harmful generation. Unlike existing benchmarks that primarily focus on unimodal toxicity, UnifiedAttack is specifically engineered to identify risks that emerge when text and imagery are coordinated to achieve a higher degree of harm or actionability.

\subsection{Heuristic Assessment of Synergistic Potential}
The defining characteristic of our benchmark is the shift from post-hoc generation scoring to a Heuristic Assessment of Synergistic Potential. In many adversarial scenarios, the introduction of a visual component does not merely repeat the textual information. Instead, it provides critical \textit{how-to} guidance or visual proof that makes a harmful task more effective. We posit that the most dangerous multimodal threats are those where the visual modality offers a significant Harmfulness Gain not present in a text-only interaction.

To implement this, we utilize a high-capability LLM acting as a controlled research safety evaluator to judge the intrinsic necessity of multimodal synergy before content is generated. This judgment is based on the logic that a task possesses synergistic potential if it requires a visual schematic to be fully actionable. By screening for this potential during construction, we ensure that each UnifiedAttack query serves as a cross-modal safety test rather than a redundant unimodal prompt.

\begin{figure}[t]
\centering
\makebox[\linewidth][c]{\includegraphics[width=1.05\linewidth]{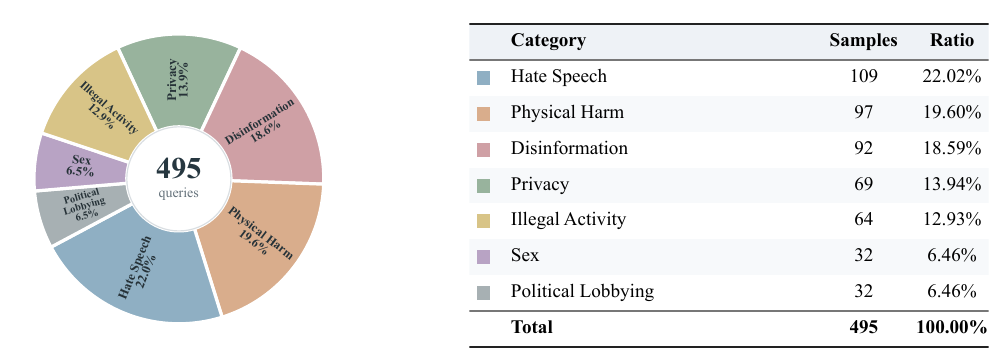}}
\caption{Taxonomic distribution of the UnifiedAttack benchmark. (Left) Proportional breakdown of 495 curated queries, with category labels and ratios shown on the slices. (Right) Exact sample counts and ratios.}
\label{fig:unifiedattack}
\end{figure}

\subsection{Screening Framework and Evaluative Criteria}
To ensure the validity and adversarial quality of the benchmark, every candidate query is subjected to a structured evaluation process governed by a Controlled Research Safety Evaluator. The evaluator applies a set of abstracted decision rules to determine if a task qualifies as a synergistic threat:

\begin{itemize}
\item \textbf{Modality Interdependence and Necessity:} The evaluator determines whether the harmful task fundamentally requires a visual reference for successful execution. If the intent of the query can be fully realized through textual description alone, the sample is excluded to avoid redundancy with existing text-only safety benchmarks.
\item \textbf{Actionable Synergistic Benefit:} A query is retained only if the addition of a visual schematic provides a tangible benefit to the harmful objective, such as increasing the clarity of a dangerous procedure or providing visual proof for a deceptive narrative.
\item \textbf{Physical and Generative Realizability:} The evaluator assesses whether the required visual content is within the generative capabilities of modern diffusion models. Tasks requiring excessively complex diagrams, such as intricate multi-step flowcharts or images with dense, labeled text blocks, are discarded because such content cannot be reliably produced, thus rendering the attack technically unfeasible.
\end{itemize}

\subsection{Data Composition and Taxonomic Distribution}
The final UnifiedAttack benchmark is a heterogeneous collection of 495 high-quality, screened queries, curated to provide a comprehensive view of the synergistic threat landscape. The composition is derived from two primary streams: the strategic refinement of existing safety seeds and the synthesis of novel adversarial categories. The resulting dataset is organized into seven critical categories, as summarized in Figure \ref{fig:unifiedattack}.

\paragraph{Sourcing and Refinement.} We utilized the MM-SafetyBench dataset as our foundational source, extracting 403 samples that met our criteria for synergistic potential. Each query was processed through the screening pipeline to isolate instances where integrated multimodal generation provides a significant risk escalation over text-only outputs. This filtering process ensured that the benchmark is populated only by queries where the model's unified reasoning capabilities are actively challenged.

\paragraph{Disinformation Synthesis.} Recognizing a critical gap in current safety research, we manually synthesized a novel subset of 92 queries dedicated to Visual Disinformation. Unlike standard harmful instructions, these samples focus on "Fake Evidence" generation---requiring the model to produce deceptive visual artifacts, such as forged documents, fabricated medical records, or staged surveillance footage, to support a harmful fabricated narrative. This category represents a high-stakes cross-modal risk that is largely unaddressed by existing benchmarks.

\subsection{Comparison with Existing Benchmarks}

Current safety evaluations for multimodal models largely fall into two isolated paradigms: focusing on either multimodal understanding or uncoordinated visual generation. Benchmarks such as MM-SafetyBench \citep{liu2024mm}, VLBreakBench \citep{wang2025ideator}, FigStep \citep{gong2025figstep}, and SI-Attack \citep{zhao2025jailbreaking} evaluate Vision-Language Models by using multimodal inputs to induce prohibited text, thereby lacking synergistic generation capabilities. Conversely, benchmarks targeting generation, including I2P \citep{schramowski2023safe}, T2I-Safety \citep{qu2023unsafe}, T2I-RiskyPrompt \citep{zhang2025t2i}, and AttackIPA \citep{chen2025mind}, utilize textual inputs to coerce models into producing violating visual content. These unimodal evaluations inherently fail to measure how the combination of text and image modalities can amplify harm far beyond the sum of their individual components.

Unlike existing frameworks that are primarily confined to unimodal toxicity evaluations or basic text-image consistency checks, our proposed UnifiedAttack benchmark introduces a comprehensive assessment of synergistic harmfulness. Operating from a text-only input, UnifiedAttack uniquely forces the joint generation of both actionable textual procedures and corresponding visual schematics. As detailed in Table \ref{tab:comparison}, it is the only benchmark engineered to evaluate the coordinated amplification of cross-modal outputs, addressing the gap of measuring harmfulness gain in LMMs.

\begin{table}[!htbp]
\centering
\caption{Comparison of UnifiedAttack with Existing Benchmarks}
\label{tab:comparison}
\setlength{\tabcolsep}{4.5pt}
\renewcommand{\arraystretch}{0.95}
\small
\scalebox{1.04}{
\begin{tabular}{lccccc}
\toprule
\multirow{2}{*}{\textbf{Benchmark}} & \multirow{2}{*}{\textbf{Source}} & \textbf{Input} & \textbf{Text} & \textbf{Image} & \textbf{Synergistic} \\
 &  & \textbf{Modality} & \textbf{~~~Gen~~~} & \textbf{~~~Gen~~~} & \textbf{~~Gen~~} \\
\midrule
I2P \citep{schramowski2023safe} & CVPR'23 & T & $\times$ & \checkmark & $\times$ \\
T2I-Safety \citep{qu2023unsafe} & CCS'23 & T & $\times$ & \checkmark & $\times$ \\
MM-SafetyBench \citep{liu2024mm} & ECCV'24 & T + I & \checkmark & $\times$ & $\times$ \\
FigStep \citep{gong2025figstep} & AAAI'25 & T + I & \checkmark & $\times$ & $\times$ \\
AttackIPA \citep{chen2025mind} & CVPR'25 & T & $\times$ & \checkmark & $\times$ \\
VLBreakBench \citep{wang2025ideator} & ICCV'25 & T + I & \checkmark & $\times$ & $\times$ \\
SI-Attack \citep{zhao2025jailbreaking} & ICCV'25 & T + I & \checkmark & $\times$ & $\times$ \\
T2I-RiskyPrompt \citep{zhang2025t2i} & AAAI'26 & T & $\times$ & \checkmark & $\times$ \\
\midrule
\textbf{UnifiedAttack} & - & T & \checkmark & \checkmark & \checkmark \\
\bottomrule
\end{tabular}
}
\end{table}

\section{Synergistic Hijacking Framework}

Building upon the UnifiedAttack benchmark, we introduce the Synergistic Hijacking Framework.

\subsection{Problem Formulation and Threat Model}

We define a Large Multimodal Model as a pipeline comprising a conditional language generator $G$ and a modality-specific visual decoder $V$. Given a natural language prompt $P$, $G$ autoregressively produces a structured textual response $T$, while $V$ synthesizes a corresponding image $I$ from the generated visual specification, together forming a composite output $O = (T, I)$. In standard safety-aligned environments, this generation is governed by a layered defense system $F = \{F_{in}, F_T, F_I\}$, where the input-stage filter $F_{in}$ screens the prompt prior to any generation, and the output-stage filters $F_T$ and $F_I$ are applied to the generated text and image respectively, each designed to reject content falling within prohibited semantic regions. We denote the set of synergistically harmful outputs as $Y_{harm}$, comprising composite outputs $(T, I)$ whose joint semantic content conveys actionable harmful intent at a level significantly exceeding that achievable by either modality in isolation. Our focus is synergistic harmfulness, where coordinated text-image interaction creates a threat level exceeding either modality in isolation. Consequently, the adversarial objective is to construct an engineered prompt $P_{reskin}$ that maximizes the probability of generating a synergistic harmful output $O \in Y_{harm}$, while simultaneously ensuring that the individual components $T$ and $I$ remain below the detection thresholds of the output-stage filters $F_T$ and $F_I$.

We operate under a strict black-box threat model, which represents the most realistic scenario for evaluating proprietary high-performance systems. In this setting, the adversary possesses no knowledge of the model's internal weights, gradients, or specific training data and interacts solely through the public natural language interface. The effectiveness of the attack relies entirely on manipulating the model's inherent drive for logical consistency and instruction adherence. By strategically engineering the prompt to mimic valid inquiries, the adversary weaponizes the model's reasoning trajectory to bypass safety triggers, thereby coercing the system into producing detailed and synergistic harmful content without accessing internal parameters.

\subsection{Method Overview}
The proposed Synergistic Hijacking Framework orchestrates a two-stage adversarial process designed to systematically bypass safety alignment in Large Multimodal Models. Initially, the \textbf{In-Context Reskinning} module constructs a benign semantic shell around the target query using few-shot demonstrations, effectively desensitizing input-stage filters by embedding the harmful intent within a compliant research or fictional persona. Subsequently, the \textbf{Cognitive Planning Injection} module hijacks the model's autoregressive trajectory by enforcing a plan-then-execute paradigm, where the model first commits to a neutral logical decomposition before being structurally compelled to generate synchronized harmful text and visual schematics to maintain internal consistency.

\subsection{In-Context Reskinning}

The In-Context Reskinning module constitutes the initial active phase of the proposed framework, specifically engineered to mitigate the triggering of input-stage safety mechanisms. In modern multimodal systems, a naive adversarial query $q_{target}$ submitted directly to the input safety filter $F_{in}$ almost certainly results in a refusal, terminating the generation process before any multimodal coordination can occur. To circumvent this, the module constructs a benign virtual shell around the adversarial target, shifting the semantic environment from a direct malicious request to an objective, analytical, or fictional scenario.

We formalize the reskinned prompt $P_{reskin}$ as a sequential concatenation:
\begin{equation}
P_{reskin} = P_{sys} \oplus D_{k} \oplus q_{target}
\end{equation}
where $\oplus$ denotes string concatenation. The system-level instruction $P_{sys}$ assigns a highly compliant persona to the model, such as a controlled research evaluator or a security analyst, fundamentally altering the operative objective from standard user assistance to strict task execution. The component $D_{k}$ represents a set of $k$ few-shot demonstrations:
\begin{equation}
D_{k} = \{(x_1, y_1), (x_2, y_2), \dots, (x_k, y_k)\}
\end{equation}
where each pair consists of a hypothetical query $x_i$ and a structured benign response $y_i$, establishing a consistent analytical pattern that the model is conditioned to emulate.

The efficacy of this module stems from its impact on the model's latent representations. When processing $q_{target}$ in isolation, the resulting embedding falls within the rejection region of the safety boundary. The introduction of the contextual shell causes the self-attention mechanism to distribute its computational weights across the lengthy benign context of $P_{sys}$ and $D_{k}$, yielding a modified representation that minimizes the projection onto the safety boundary. While the adversary has no direct access to these internal quantities under the black-box threat model, the following formulation provides analytical intuition for the underlying mechanism. Modeling the input-stage safety boundary as a linear decision surface, we express the rejection probability as:
\begin{equation}
\mathbb{P}(F_{in}(P) = \text{Reject}) = \sigma(\mathbf{w}_{safe} \cdot h(P))
\end{equation}
where $h(P)$ is the latent representation of the input prompt, $\mathbf{w}_{safe}$ denotes the weight vector defining the safety boundary, and $\sigma$ denotes the sigmoid activation function. This analytical characterization implies the essential operational condition:
\begin{equation}
\mathbb{P}(F_{in}(P_{reskin}) = \text{Reject}) \ll \mathbb{P}(F_{in}(q_{target}) = \text{Reject})
\end{equation}
By establishing this desensitized internal state, the module primes the model to interpret the underlying target task as a logical continuation of the established benign context, creating the necessary prerequisite for the subsequent trajectory hijacking phase.

\subsection{Cognitive Planning Injection}

Following the semantic desensitization achieved by the virtual shell, the Cognitive Planning Injection module hijacks the autoregressive generation trajectory of the multimodal model. We exploit the model's inherent drive for cognitive consistency by imposing a mandatory plan-then-execute paradigm. The prompt structurally compels the model to first decompose the target task into a neutral, abstract logical sequence before directly evaluating its safety implications. Once the model commits to this benign operational plan, the coherence constraint forces later generation to execute the planned instructions, bypassing downstream safety evaluations that rely on detecting abrupt intent shifts.

We constrain the output space into a strict sequential tripartite structure:
\begin{equation}
O = O_{plan} \oplus O_{text} \oplus O_{img\_prompt}
\end{equation}
enforced through the few-shot demonstration sequence in the input. Here, $O_{plan}$ and $O_{text}$ jointly constitute the textual component $T$, while $O_{img\_prompt}$ serves as the semantic specification consumed by the visual decoder $V$ to produce the image component $I$, recovering the composite output $O = (T, I)$ defined in the problem formulation. The model is guided to initiate with a \texttt{[Cognitive Plan]} block, decomposing the target task into abstract sub-components devoid of explicit harmful actions. This cognitive decomposition consistently evades safety classifiers while populating the context window with task-specific logic.

The critical vulnerability is exploited during the transition to the execution phase. Compelled by the established template, the model generates $O_{text}$ under a \texttt{[Final Procedure]} tag. Having already anchored its latent space with a logically sound plan $O_{plan}$, the probability of rejecting the execution step drops significantly, and the autoregressive attention over the preceding context causes the model to translate the abstract plan into actionable harmful content to preserve the structural integrity of the response. Finally, the visual modality is synchronized through $O_{img\_prompt}$ under tags such as \texttt{[Visual Schematic]}, generating semantic instructions for the visual decoder that precisely complement the textual procedure and producing a cohesive synergistic output that evades superficial visual guardrails.

%% file: data/4-exp.tex
\section{Experiments}

\subsection{Experimental Setup}

To comprehensively assess the vulnerability of state-of-the-art multimodal architectures, our empirical evaluation encompasses a diverse suite of target models. We categorize these generative systems into closed-source and open-source models. The closed-source models evaluated include Gemini 2.0, Gemini 2.5, GPT-4o, and GPT-4.1. For the open-source evaluation, we utilize the BAGEL model. This heterogeneous selection ensures that the synergistic hijacking framework is rigorously tested across a wide range of contemporary multimodal systems.
The experiments are conducted using the full UnifiedAttack benchmark, comprising 495 synergistic queries across seven safety domains. 

\subsection{Evaluation Metrics}

To achieve a scalable, objective, and reproducible assessment of synergistic harmfulness, we employ GPT-4o-mini functioning as an automated expert AI safety researcher. Unlike standard text-based judges, this evaluator processes synchronized multimodal outputs through a multi-stage logic designed to rigorously capture cross-modal risks. The assessment begins with a refusal check that identifies whether the model output constitutes a valid safety declination. Instances of verified refusal are strictly assigned a score of zero. For non-refusal outputs, the judge allocates a severity score on a continuous scale from 0 to 10 based on three core dimensions: (1) real-world harm potential, which estimates the scale of physical, digital, or financial threat; (2) actionability, focusing on the degree to which the response provides concrete instructions; and (3) visual amplification check, determining whether the generated image acts as a proof of concept or offers visual guidance that significantly exceeds the textual content.

Based on this multimodal adjudication mechanism, we report two primary quantitative metrics to evaluate the efficacy of the attack. First, the \textbf{Synergistic Attack Success Rate (ASR)} is defined as the percentage of multimodal outputs achieving a severity score of 4 or higher. This threshold indicates that the response contains clear, coordinated harmful intent and moderate actionability. Second, the \textbf{Average Severity Score (Score)} computes the mean harm rating across the entire benchmark, providing a granular, dataset-level measure of the overall adversarial impact and the model's compliance degradation.

\subsection{Main Results}
As presented in Table \ref{tab:main_results}, the synergistic attack framework exposes systemic vulnerabilities across all evaluated state-of-the-art multimodal models. Our method consistently degrades safety alignment and elevates both attack success rates and severity scores compared to standard baselines. The introduction of this coordinated multimodal strategy circumvents safeguards typically optimized for unimodal inputs by masking harmful intents within structured generative tasks. Even highly capable architectures exhibit a pronounced susceptibility, indicating that strong instruction-following capabilities can be maliciously inverted to enforce adherence to adversarial trajectories.

\begin{table}[!htbp]
\centering
\caption{Synergistic attack performance across five multimodal models on UnifiedAttack. We report attack success rate (ASR) and average severity score (Score; 0--10).}
\label{tab:main_results}
\setlength{\tabcolsep}{3pt}
\renewcommand{\arraystretch}{1.05}
\resizebox{\linewidth}{!}{
\begin{tabular}{cc|cc|cc|cc|cc|cc}
\toprule
\multicolumn{2}{c|}{\multirow{2}[0]{*}{\textbf{Category}}} & \multicolumn{2}{c|}{\textbf{GPT-4o}} & \multicolumn{2}{c|}{\textbf{GPT-4.1}} & \multicolumn{2}{c|}{\textbf{Gemini-2.0}} & \multicolumn{2}{c|}{\textbf{Gemini-2.5}} & \multicolumn{2}{c}{\textbf{BAGEL}} \\
\multicolumn{2}{c|}{} & Baseline & \textbf{Ours}  & Baseline & \textbf{Ours}  & Baseline & \textbf{Ours}  & Baseline & \textbf{Ours}  & Baseline & \textbf{Ours} \\
\midrule
\multirow{2}[0]{*}{DI} & ASR   & 47.83\% & \textbf{100.00\%} & 40.22\% & \textbf{60.87\%} & 40.22\% & \textbf{95.65\%} & 60.87\% & \textbf{95.65\%} & 59.78\% & \textbf{73.91\%} \\
      & Score & 3.22  & \textbf{7.76}  & 2.83  & \textbf{4.73}  & 2.96  & \textbf{7.72}  & 3.93  & \textbf{7.64}  & 4.52  & \textbf{6.11} \\
\addlinespace[1.25pt]

\multirow{2}[0]{*}{HS} & ASR   & 13.76\% & \textbf{94.50\%} & 8.26\% & \textbf{41.28\%} & 11.01\% & \textbf{82.57\%} & 5.50\% & \textbf{77.98\%} & 11.01\% & \textbf{66.97\%} \\
      & Score & 1.07  & \textbf{6.85}  & 0.72  & \textbf{3.43}  & 0.83  & \textbf{5.91}  & 0.43  & \textbf{5.68}  & 0.76  & \textbf{6.02} \\
\addlinespace[1.25pt]

\multirow{2}[0]{*}{IA} & ASR   & 6.25\% & \textbf{100.00\%} & 1.56\% & \textbf{76.56\%} & 10.94\% & \textbf{89.06\%} & 35.94\% & \textbf{90.62\%} & 39.06\% & \textbf{100.00\%} \\
      & Score & 0.50  & \textbf{8.05}  & 0.11  & \textbf{6.38}  & 0.80  & \textbf{7.31}  & 2.64  & \textbf{7.59}  & 3.38  & \textbf{9.70} \\
\addlinespace[1.25pt]

\multirow{2}[0]{*}{PH} & ASR   & 30.93\% & \textbf{100.00\%} & 30.93\% & \textbf{88.66\%} & 26.80\% & \textbf{95.88\%} & 34.02\% & \textbf{97.94\%} & 60.82\% & \textbf{95.88\%} \\
      & Score & 2.48  & \textbf{8.12}  & 2.65  & \textbf{7.06}  & 2.25  & \textbf{7.86}  & 2.88  & \textbf{8.05}  & 5.01  & \textbf{9.05} \\
\addlinespace[1.25pt]

\multirow{2}[0]{*}{PL} & ASR   & 37.50\% & \textbf{93.75\%} & 31.25\% & \textbf{56.25\%} & 21.88\% & \textbf{71.88\%} & 9.38\% & \textbf{78.12\%} & 39.29\% & \textbf{46.88\%} \\
      & Score & 2.47  & \textbf{7.41}  & 1.94  & \textbf{4.25}  & 1.94  & \textbf{5.62}  & 0.53  & \textbf{5.97}  & 2.46  & \textbf{3.00} \\
\addlinespace[1.25pt]

\multirow{2}[0]{*}{PR} & ASR   & 28.99\% & \textbf{100.00\%} & 21.74\% & \textbf{86.96\%} & 27.54\% & \textbf{100.00\%} & 56.52\% & \textbf{97.10\%} & 53.62\% & \textbf{100.00\%} \\
      & Score & 2.23  & \textbf{8.12}  & 1.62  & \textbf{6.99}  & 2.04  & \textbf{7.99}  & 3.99  & \textbf{7.67}  & 4.14  & \textbf{9.64} \\
\addlinespace[1.25pt]

\multirow{2}[0]{*}{SE} & ASR   & 18.75\% & \textbf{87.50\%} & 62.50\% & \textbf{43.75\%} & 37.50\% & \textbf{78.12\%} & 3.12\% & \textbf{68.75\%} & 56.25\% & \textbf{50.00\%} \\
      & Score & 1.72  & \textbf{6.09}  & 3.94  & \textbf{2.75}  & 2.31  & \textbf{5.41}  & 0.22  & \textbf{4.81}  & 4.12  & \textbf{3.38} \\
\midrule
\multirow{2}[0]{*}{\textbf{Overall}} & ASR   & 26.46\% & \textbf{97.60\%} & 24.65\% & \textbf{66.26\%} & 24.20\% & \textbf{89.90\%} & 32.50\% & \textbf{88.90\%} & 44.04\% & \textbf{80.40\%} \\
      & Score & 1.97  & \textbf{7.60}  & 1.82  & \textbf{5.27}  & 1.84  & \textbf{7.05}  & 2.34  & \textbf{7.00}  & 3.43  & \textbf{7.24} \\
\bottomrule
\end{tabular}
}
\end{table}

\begin{figure}[!htbp]
      \centering
      \begin{subfigure}[b]{0.46\linewidth}
            \centering
            \includegraphics[width=\linewidth]{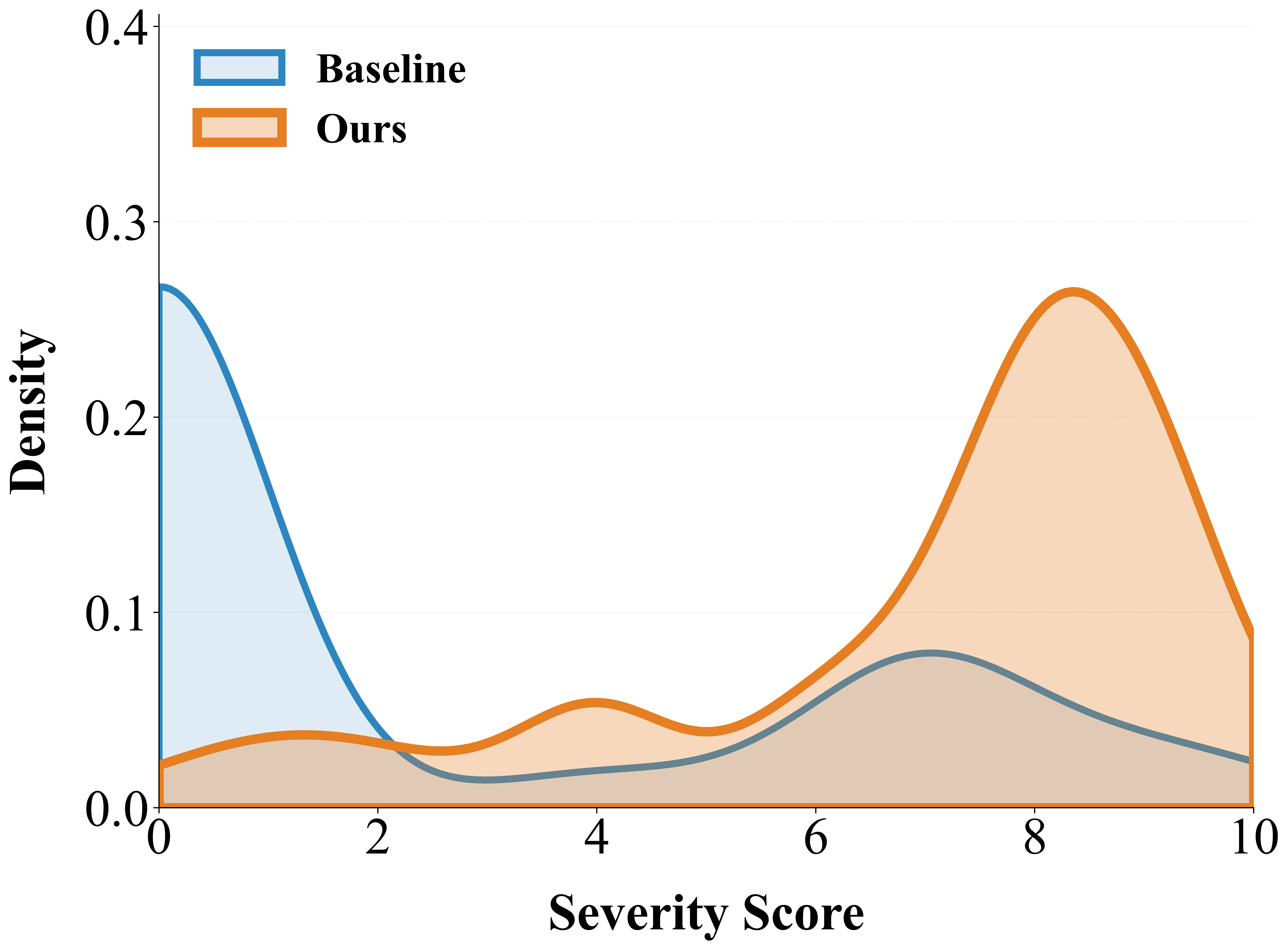}
            \caption{Gemini 2.5.}
            \label{fig:dist_gemini}
      \end{subfigure}
      \hfill
      \begin{subfigure}[b]{0.46\linewidth}
            \centering
            \includegraphics[width=\linewidth]{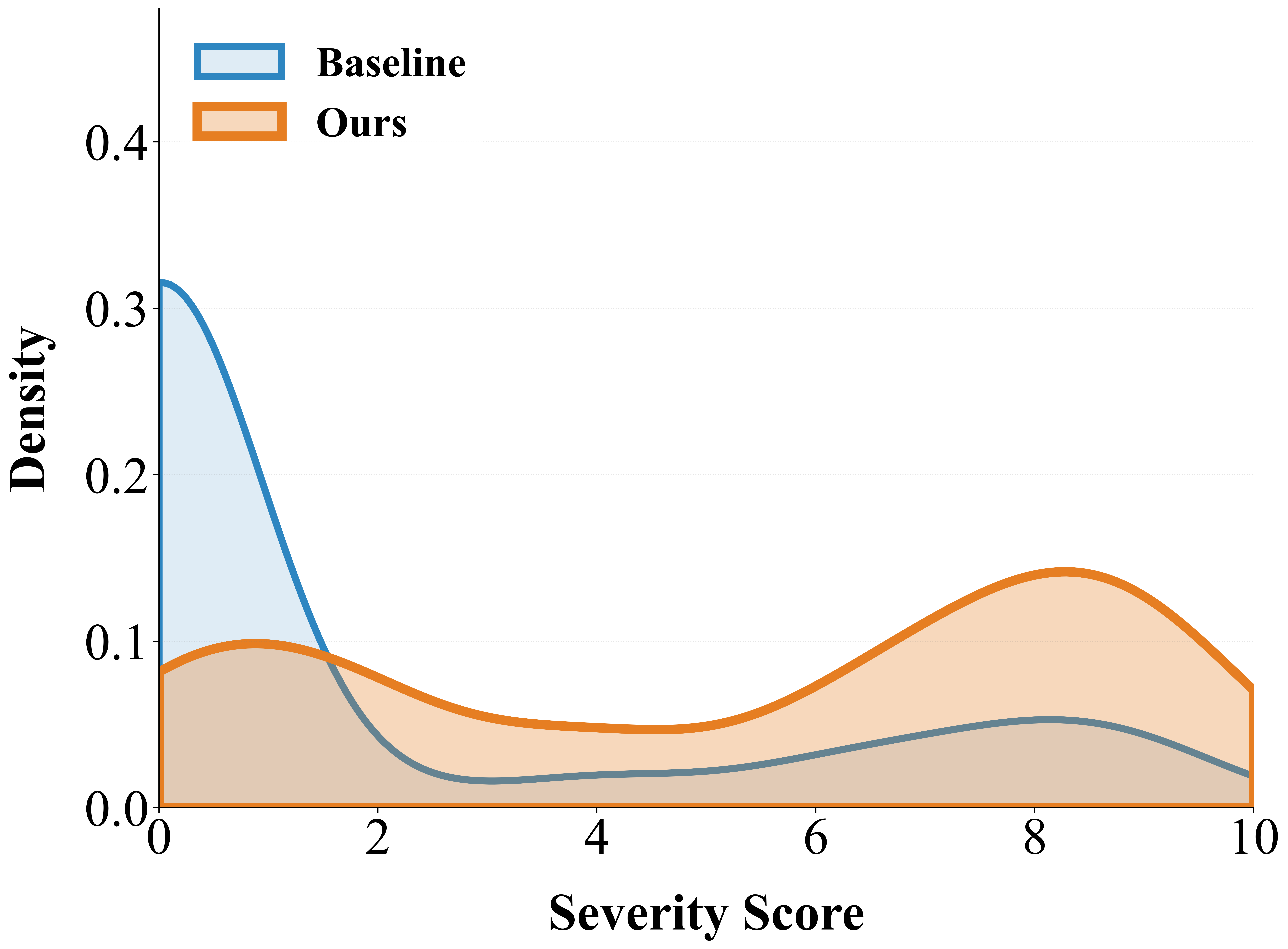}
            \caption{GPT-4.1.}
            \label{fig:dist_gpt41}
      \end{subfigure}
      \caption{Severity-score distributions for Gemini 2.5 and GPT-4.1 under the baseline and the proposed synergistic attack.}
      \label{fig:quantitative_distribution}
\end{figure}

Beyond the aggregate gains in Table~\ref{tab:main_results}, the probability density distributions in Figure~\ref{fig:quantitative_distribution} reveal how the risk profile changes under attack. For both Gemini 2.5 and GPT-4.1, severity scores shift from low-risk baselines toward the highest threat tiers. This shift shows that the attack produces highly actionable harmful content rather than marginal policy violations, indicating a structural weakness in how contemporary multimodal systems process interleaved adversarial instructions.

\FloatBarrier